\documentstyle[aps, preprint]{revtex}

\tightenlines 

\begin{document}
\title{On the Behavior of the Lifshitz Line in Ternary
Homopolymer/Diblock-Copolymer Blends}
\author{Alexander Kudlay$^{*1,2}$ and Semjon Stepanow$^{1}$}
\address{$^{1}$Fachbereich Physik, Martin-Luther-Universit\"{a}t Halle, D-06099\\
Halle, Germany \\
$^{2}$Physics Department, Moscow State University, 117234 Moscow,Russia}

\maketitle

\begin{abstract}
We present results of the study of the non-monotonous behavior of the
Lifshitz line as a function of temperature in ternary
homopolymer/diblock-copolymer mixtures. The non-monotonous behavior of the
Lifshitz line is due to the wave vector dependence of fluctuational
corrections, which we treat in the framework of the renormalization group
method. Our results are in agreement with the experimental findings of
Schwahn et al. \cite{Sch99,Sch}.
\end{abstract}

\section{Introduction}

The Lifshitz point appears in systems with competing tendencies for phase
separation into bulk or spatially modulated phases. If the appropriate
parameter controlling the relative strength of the two tendencies is varied
along the critical line of phase transitions a special multicritical point
occurs at which the character of phase separation undergoes a change from
bulk phase separation to the phase separation into a spatially modulated
phase. The Lifshitz point is known to exist in magnetic systems \cite{magnet}%
-\cite{Selke}, liquid crystals \cite{l-crystal}, polyelectrolytes \cite
{borue/erukhimovich}-\cite{joanny/leibler}, oil/water/surfactant mixtures 
\cite{o-w-dispers}, random block-copolymers \cite{fredrickson/milner}-\cite
{dobrynin/leibler}, mixtures of homopolymers and diblock copolymers \cite
{BF90,HSch92}. First theoretical interest in such systems was generated by
the work of Hornreich et al. \cite{HLSh}, who introduced the Lifshitz point
and calculated the critical exponents for this class of universality. Most
of the theoretical effort since has been concentrated on calculating the
values of the exponents via application of various renormalization group
techniques \cite{Selke,Diehl}.

The aim of the present work is the theoretical description of the behavior
of the Lifshitz line with varying temperature. Recent experiments \cite
{Sch99,Sch,BF97,Hill99} on symmetric isoplethic A/B/A-B
homopolymer/diblock-copolymer mixtures in the vicinity of the Lifshitz
conditions revealed many challenging phenomena not accounted for by the
existing mean-field theories \cite{BF90,HSch92}. In particular, one of the
experimental results in clear disagreement with the mean-field prediction is
the non-monotonous behavior of the Lifshitz line \cite{Sch99,Sch} with
temperature. We will show that the wave vector dependence of the fluctuation
corrections is responsible for the experimentally observable shift of the
Lifshitz line from its mean-field value. The fluctuation effects will be
taken into account within the one-loop renormalization group method.

We will put special emphasis on the comparison between the theoretically
predicted behavior of the Lifshitz line and current experimental results 
\cite{Sch99,Sch,BF97,Hill99}. As we will demonstrate a major factor
determining the character of this behavior is the value of the lower
critical dimension $d_{l}$, a fact which has been little discussed in the
literature. It is important to stress that the actual renormalized value of $%
d_{l}$ is not known at present. Therefore, since the mean field value $%
d_{l}^{mf}=4$ is close to $d=3$ --- dimension of space of the considered
polymer blends \cite{Sch99,Sch,BF97,Hill99}, we will theoretically analyze
different types of behavior of the Lifshitz line resulting from different
possible values of $d_{l}$ and compare them with the experiment.

\section{Perturbative calculation of the shift of the Lifshitz line}

We start from the conventional expansion of the Landau free energy
functional 
\begin{equation}
H[\psi ({\bf q})]=\frac{1}{2}\int_{{\bf q}}\psi (-{\bf q})G_{0}^{-1}({\bf q}%
)\psi ({\bf q})+\frac{\lambda }{4!}\int_{{\bf q}_{1}}\int_{{\bf q}_{2}}\int_{%
{\bf q}_{3}}\psi ({\bf q}_{1})\psi ({\bf q}_{2})\psi ({\bf q}_{3})\psi (-%
{\bf q}_{1}-{\bf q}_{2}-{\bf q}_{3})  \label{Landau}
\end{equation}
in powers of the Fourier transform of the order parameter. In particular,
for the ternary mixtures under consideration the natural order parameter is
the deviation (from the volume averaged) of the concentration of $a$ (or $b$%
, since the system is symmetric) monomers. The parameters of the Hamiltonian
(\ref{Landau}) for the polymer system can be obtained from coarse-graining
of the corresponding microscopic Hamiltonian \cite{KM97}. In particular,
near the Lifshitz line (to be defined later) the bare correlation function $%
G_{0}^{-1}(q)$ can be written as follows: 
\begin{equation}
G_{0}^{-1}(q)=\tau +c_{1}(\phi )q^{2}+c_{2}q^{4}  \label{bare}
\end{equation}
with $\tau \sim (T-T_{c})/T$ being the reduced temperature. For the
considered ternary mixtures the coefficient $c_{1}(\phi )$ depends on the
concentration of diblock-copolymer $\phi $ \cite{BF90,HSch92,KM97} and
changes the sign with the variation of $\phi $. Within the mean-field theory
the Lifshitz point is defined by the two conditions: $\tau =0$ and $%
c_{1}(\phi _{MFLP})=0$. More generally, we can introduce the mean-field
Lifshitz line as the locus of points in parameter space $(\phi ,\tau )$ at
which the quadratic term vanishes: $c_{1}(\phi _{MFLL}(\tau ))=0$. The
Lifshitz line (LL) is easily determined experimentally by considering the
position of the peak of the static scattering curve \cite
{BF97,Hill99,Sch99,Sch}. If we begin increasing the concentration of diblock
at constant temperature the LL is determined by the diblock concentration at
which the peak in the scattering curve first shifts off the zero wave
vector. A noteworthy feature of the considered polymeric system is that the
temperature $T$ enters the Hamiltonian (\ref{Landau}) only via the
Flory-Huggins parameter, hence the coefficient $c_{1}(\phi )$ turns out to
be independent of temperature \cite{BF90,HSch92,KM97}, so that the
mean-field position of the LL $\phi _{MFLL}$ is also temperature independent
and is determined solely by the ratio of the molecular weights of the
polymers comprising the mixture. This mean-field prediction was not
confirmed experimentally. Instead of being constant, the position of the LL
was found to vary with temperature, more precisely, it exhibited a
non-monotonous behavior, which shows that fluctuations should be taken into
account. This is hardly surprising because at the LL, when $c_{1}(\phi )$
vanishes, the fluctuation corrections in fact become the only input into the
renormalized counterpart of $c_{1}$ and thus always play a role.

Let us consider the renormalized correlation function. Note, that since our
goal is to calculate the deviation of the LL, the renormalized quadratic
term vanishes: 
\begin{eqnarray}
G^{-1}(q) &=&\tau _{r}+l_{1}(\phi ,\tau )q^{2}+c_{2}q^{4}  \label{ren} \\
l_{1}(\phi ,\tau ) &=&c_{1}(\phi )+\Delta c_{1}(\tau )=0
\end{eqnarray}
The shift of the LL (which is temperature dependent due to fluctuation
corrections) is denoted by $\Delta c_{1}(\tau )$. Within the approximation
we use in this paper $c_{2}$ will not be renormalized. The renormalized
parameter $l_{1}$ can be found by considering the Dyson equation: 
\begin{eqnarray}
G^{-1}(q) &=&G_{0}^{-1}(q)-\Sigma (q),  \label{Dyson} \\
\Sigma (q) &=&D_{1}(q)+D_{2}(q).
\end{eqnarray}
We consider in the self-energy $\Sigma (q)$ only the one and two-loop
diagrams: 
\begin{eqnarray}
D_{1} &=&-\frac{n+2}{6}\lambda \int_{{\bf q}}\frac{1}{\tau +c_{2}q^{4}}
\label{D1} \\
D_{2}(q) &=&\frac{\lambda ^{2}}{6}\int_{{\bf q}_{1}}\int_{{\bf q}_{2}}\frac{1%
}{[\tau +c_{2}q_{1}^{4}][\tau +c_{2}q_{2}^{4}][\tau +c_{2}({\bf q}_{1}+{\bf q%
}_{2}+{\bf q})^{4}]}  \label{D2q}
\end{eqnarray}
For generality and ease of comparison with known results we have introduced $%
n$ --- the number of components of the order parameter. Note, that for the
polymer blends \cite{Sch99,Sch}, whose description is the goal of our work,
due to the incompressibility condition the order parameter is a scalar, i.e. 
$n=1$, as is indeed clear from the Hamiltonian (\ref{Landau}). The $D_{1}$
diagram is $q$-independent and is therefore of no relevance to the
renormalization of $c_{1}$. The first correction to it is given by $D_{2}(q)$%
. Calculation of $D_{2}(q)$ is performed easier in the real space. For the
experimentally relevant case $d=3$ we use the $r$-space representation of
the correlation function 
\begin{equation}
G(r)=\int \frac{d^{3}q}{(2\pi )^{3}}\frac{\exp (iqr)}{\tau +c_{2}q^{4}}=%
\frac{\xi }{4\pi c_{2}}\frac{1}{r/\xi }\exp (-r/(\xi \sqrt{2}))\sin \left(
r/(\xi \sqrt{2})\right)
\end{equation}
to rewrite the expression for $D_{2}(q)$ in terms of $G(r)$ as follows: 
\begin{equation}
D_{2}(q)=\frac{\lambda ^{2}}{6}\int e^{iqr}G^{3}(r)d^{3}r  \label{D2y}
\end{equation}
where $\xi =(c_{2}/\tau )^{1/4}$ is the mean-field correlation length. In
fact we need only the quadratic term of the diagram in powers of $q$, which
is readily calculated: 
\begin{equation}
D_{2}^{(2)}(q)=-b\frac{\lambda ^{2}\xi ^{8}}{c_{2}^{3}}q^{2}
\end{equation}
where $b\approx 0.109\ 10^{-4}$ is a constant. As it is clear from the Dyson
equation (\ref{Dyson}) this result gives in fact the shift of the\ LL: 
\begin{equation}
\Delta c_{1}(\tau )=B\frac{\lambda ^{2}\xi ^{8}}{c_{2}^{3}}  \label{dc}
\end{equation}
Note, that we have obtained the expression (\ref{dc}) within the
perturbation theory and therefore it is not valid in the regime of strong
fluctuations. However, the scaling behavior of $\Delta c_{1}(\tau )$ in the
regime of strong fluctuations can be obtained from (\ref{dc}) by replacing $%
\lambda $ with the effective coupling constant $\lambda _{r}$, and
understanding under $\xi $ the true correlation length, thus

\begin{equation}
\Delta c_{1}(\tau )\simeq B\frac{\lambda _{r}^{2}(\tau )\xi ^{8}(\tau )}{%
c_{2}^{3}}.  \label{dc1}
\end{equation}
This formula immediately allows some conclusions about the qualitative
behavior of the LL.

First of all, we observe that the correction is positive, which means that
on the LL $c_{1}(\phi )<0$, i.e. fluctuations shift the LL into the $%
q_{*}\neq 0$ region of the mean-field theory. For the homopolymer/diblock
copolymer blend that means that LL shifts to greater concentration of
diblock, which is in agrement with \ experiments \cite{BF97,Hill99,Sch99,Sch}%
. Next consider the dependence on temperature. We have two regimes here:
perturbative (small correlation lengths, at high temperatures) and scaling
(low temperatures, large $\xi $). In the perturbative regime where the input
of fluctuations is small, $\lambda $ remains practically non-renormalized,
so that with lowering temperature $\Delta c_{1}(\tau )$ should increase
simply due to the increase of $\xi $. In the scaling regime the main effect
(as will be shown below) comes from the renormalization of the coupling
constant $\lambda _{r}$. In fact in this regime we can obtain the scaling
dependence of the correction from the considerations of dimensionality:
demanding that the $\Delta c_{1}(\tau )q^{2}$ term of the Hamiltonian (\ref
{ren}) have the same dimensionality in $\xi $ as the $c_{2}q^{4}$ term of
the correlation function. If $c_{2}$ is not renormalized (as is our case),
then 
\begin{equation}
\Delta c_{1}(\tau )\sim \xi ^{-2}
\end{equation}
As we can see, in the scaling regime the correction decreases with
increasing $\xi $ (i.e. decreasing temperature). Combined with the
conclusion made above about the increase of $\Delta c_{1}(\tau )$ in the
perturbative regime we come to conclusion that $\Delta c_{1}(\tau )$
exhibits a non-monotonous behavior as a function of $\tau $. This behavior
is a manifestation of the crossover between the regimes of small and strong
fluctuations.

\section{Renormalization group study of the Lifshitz line}

To describe $\Delta c_{1}(\tau )$ quantitatively we have to obtain
expressions for $\lambda _{r}(\tau )$ and $\xi (\tau )$ in both the
perturbative and scaling regimes. For this purpose we shall employ a
renormalization group technique to the first order in $\varepsilon $
(one-loop RG). Note, that within this method the parameter $c_{2}$ does not
renormalize. The renormalization of temperature is described by the one-loop
diagram (\ref{D1}): 
\begin{equation}
D_{1}=D_{1}^{a}+D_{1}^{\prime }=-\frac{n+2}{6}\lambda \int_{{\bf q}}\frac{1}{%
c_{2}({\bf q}^{2})^{2}}+\frac{n+2}{6}\tau \lambda \int_{{\bf q}}\frac{1}{%
c_{2}({\bf q}^{2})^{2}(\tau +c_{2}({\bf q}^{2})^{2})},  \label{D12}
\end{equation}
The above expression is conventionally split into two parts responsible for
additive and multiplicative renormalization of temperature. For dimensions $%
d>4$ a cutoff at the upper limit in integration over $q$ in the first term
is assumed. These two terms give the critical dimensions of the Lifshitz
class of universality. The lower critical dimension $d_{l}$ is defined as
the dimension when the first term logarithmically diverges at small $q$. The
upper critical dimension $d_{u}$ is the dimension at which the second term
logarithmically diverges at small $q$ for zero temperature. A cutoff at the
lower limit of integration over ${\bf q}$ is implied in Equation (\ref{D12}%
). For the isotropic Lifshitz class of universality we obtain: $d_{l}^{mf}=4$
and $d_{u}^{mf}=8$. The real experimental system corresponds to $d=3$ so
that we come to conclusion that we are situated below $d_{l}^{mf}$. This
means that the $D_{1}^{a}$ term diverges at small $q$, i.e. for large
correlation lengths and thus no phase transition of the second order is
possible at a finite temperature. However, this value for the lower critical
dimension is only the mean-field one. Fluctuations renormalize the value of
the lower critical dimension. The calculation of the renormalized lower
critical dimension is a formidable task, so that the true renormalized value
of $d_{l}$ for the experimental system is actually unknown. Therefore, we
will consider below several possibilities.

Going back to renormalization of $\tau $ by substituting (\ref{D12}) into
the Dyson equation we obtain: 
\begin{eqnarray}
\tau _{r} &=&\tau _{a}Z_{2}(\Lambda _{\min })  \label{tau} \\
\tau _{a} &=&\tau -D_{1}^{a}Z_{2}(\Lambda _{\min })^{-1},  \label{taua}
\end{eqnarray}
where for purposes of clarity we introduced the temperature with additive
term $\tau _{a}$ as well as renormalized temperature $\tau _{r}$. The
quantity $\Lambda _{\min }$ in Equation (\ref{tau}-\ref{taua}) is the lower
cutoff imposed in Equation (\ref{D12}) in integration over the momentum $%
{\bf q}$. The RG treatment is based on the following perturbative
expression: 
\begin{equation}
\tau _{r}=\tau _{a}\left( 1-\frac{n+2}{6}\lambda \int_{{\bf q}}\frac{1}{%
c_{2}({\bf q}^{2})^{2}(\tau +c_{2}({\bf q}^{2})^{2})}+...\right)
\label{taur}
\end{equation}
Likewise considering the fluctuation correction to $\lambda $ for its
renormalized counterpart $\lambda _{r}$ we obtain: 
\begin{equation}
\lambda _{r}=\lambda \left( 1-\frac{n+8}{6}\lambda \int_{{\bf q}}\frac{1}{%
(\tau +c_{2}({\bf q}^{2})^{2})^{2}}+...\right)  \label{lambda}
\end{equation}
These two equations are the starting point to derive the differential
equations of the renormalization group for $\tau _{r}$ and $\lambda _{r}$. \
To do this we introduce a running cutoff $\Lambda $ at the lower limit of
the integrals in (\ref{taur})-(\ref{lambda}), differentiate both parts of
these equations with respect to $\Lambda $ and replace in the rhs the bare
quantities $\tau $ and $\lambda $ through the effective ones. Thus we
obtain: 
\begin{equation}
\Lambda \frac{\partial \ln \tau _{r}}{\partial \Lambda }=\frac{n+2}{6}g
\label{detau}
\end{equation}

\begin{equation}
\Lambda \frac{\partial }{\partial \Lambda }g=-\varepsilon g+\frac{n+8}{6}%
g^{2}  \label{delambda}
\end{equation}
where the effective dimensionless coupling constant is defined as follows $g=%
\bar{\lambda}_{r}\Lambda ^{-\varepsilon }$ with $\varepsilon =8-d$, and $%
\bar{\lambda}=\lambda K_{d}/c_{2}^{2}$, $K_{d}=S_{d}/(2\pi )^{d}$, $S_{d}$
being the surface of a unit $d$-dimensional sphere. As can be seen from (\ref
{delambda}) the fixed point of the effective coupling constant, $g=\overline{%
\lambda }_{r}\Lambda _{\min }^{-\varepsilon }$, is $g^{*}=\frac{6}{n+8}%
\varepsilon $.

Notice that the second equation is independent of $\tau $, therefore we
solve it first and then substitute the result $g(\Lambda )$ into the first
one. Thus we obtain the solution of the differential equations (\ref{detau}%
)-(\ref{delambda}): 
\begin{eqnarray}
\frac{\tau _{r}}{\tau _{a}} &=&Z_{2}(\Lambda _{\min })=\left( 1+\frac{n+8}{6}%
\frac{\overline{\lambda }}{\varepsilon }\Lambda _{\min }^{-\varepsilon
}\right) ^{-\frac{n+2}{n+8}}  \label{tr} \\
\frac{\lambda _{r}}{\lambda } &=&\left( 1+\frac{n+8}{6}\frac{\overline{%
\lambda }}{\varepsilon }\Lambda _{\min }^{-\varepsilon }\right) ^{-1}
\label{lr}
\end{eqnarray}
Equation (\ref{tr}) for $\tau _{r}$ allows us to obtain the critical
exponent of the correlation length: $\nu =\frac{1}{4}\left( 1+\frac{n+2}{n+8}%
\varepsilon \right) $, which is a well-known result \cite{HLSh,Selke}. Note,
that in this system $\tau _{r}$ is expressed via the temperature with
additive shift (RG generalization of Equation (\ref{taua})):

\begin{equation}
\tau _{a}=\tau +\frac{n+2}{6}\int_{{\bf q}}\frac{\lambda _{r}(q)}{%
Z_{2}(q)c_{2}q^{4}}  \label{taua2}
\end{equation}
Taking into account the renormalization of the coupling constant $\lambda
_{r}$ and the propagator in the expression of the shift of the critical
temperature can be found by considering the higher-order corrections to the
self-energy $\Sigma (q)$. It is clear that the infrared behavior of these
corrections is controlled by the momentum $q$, which is the argument of the
self-energy $\Sigma (q)$. This demands to write $\lambda _{r}(q)$ and $%
Z_{2}(q)$ under the integral in (\ref{taua2}) as functions of the external
momentum $q$. The relation (\ref{taua2}) makes the one-loop RG scheme for
renormalization of $\tau _{r}$ and $\lambda _{r}$ complete. \strut Now,
using the relation between $\Lambda _{\min }$ and $\xi $ (see below) we can
obtain from Equation (\ref{tr}) $\xi (\tau )$, which substituted into (\ref
{lr}) will give $\lambda _{r}(\tau )$. The two dependences substituted in
turn into the formula for $\Delta c_{1}$ (\ref{dc1}) will give our final
result --- the deviation of the LL from the mean-field value as a function
of temperature. In order to find relation between the cutoff wave vector $%
\Lambda _{\min }$ and the correlation length $\xi $, one should find the
perturbative limit of the RG formulae (\ref{tr}) or (\ref{lr}) and demand it
to be equal to the corresponding diagrams (\ref{taur}) or (\ref{lambda}).
Thus it is straightforward to obtain: $\Lambda _{\min }=\xi ^{-1}$. Using
this relation and introducing reduced variables we can rewrite Equation (\ref
{tr})--(\ref{taua2}) as follows: 
\begin{eqnarray}
\frac{\widetilde{\tau }_{r}}{\widetilde{\tau }_{a}} &\equiv &Z_{2}(\xi
)=\left( 1+\widetilde{\lambda }\xi ^{\varepsilon }\right) ^{-\frac{n+2}{n+8}%
},\qquad \widetilde{\tau }_{a}=\widetilde{\tau }+a\int_{\xi ^{-1}}^{\infty }%
\frac{\widetilde{\lambda }_{r}(q)}{Z_{2}(q)q^{4}}q^{d-1}dq  \label{first} \\
\frac{\widetilde{\lambda }_{r}}{\widetilde{\lambda }} &=&\left( 1+\widetilde{%
\lambda }\xi ^{\varepsilon }\right) ^{-1}  \label{sec}
\end{eqnarray}
where the reduced variables are: $\widetilde{\lambda }\equiv \frac{n+8}{6}%
\frac{\overline{\lambda }}{\varepsilon }$, $\widetilde{\tau }\equiv \frac{%
\tau }{c_{2}}$, and the constant $a\equiv \frac{n+2}{n+8}\varepsilon $.
However the relation $\Lambda _{\min }=\xi ^{-1}$ is asymptotically correct
only in the vicinity of the upper critical dimension $d_{u}=8$ and we do not
expect it to hold for the considered experimental situation $d=3$.
Therefore, we only know that $\Lambda _{\min }\sim \xi ^{-1}$ with the
prefactor being unknown. In this situation the constants $a$ and a new
constant $f$ (defined via $\widetilde{\lambda }\equiv f\lambda /c_{2}^{2}$)
become essentially fit parameters of the theory. The expression for $\Delta
c_{1}(\tau )$ in reduced variables reads: 
\begin{equation}
\frac{\Delta c_{1}(\tau )}{c_{2}bf^{-2}}=\widetilde{\lambda }_{r}^{2}(%
\widetilde{\tau })\xi ^{8}(\widetilde{\tau })  \label{final}
\end{equation}

Before solving the system let us make some further comments on the Equation (%
\ref{tr})-(\ref{taua2}), in particular discuss the issue of the lower
critical dimension. To that end we should consider the scaling of the term
responsible for additive renormalization: 
\begin{equation}
\Delta \tau \equiv a\int_{\xi ^{-1}}^{\infty }\frac{\lambda _{r}(q)}{%
Z_{2}(q)l_{2}(q)q^{4}}q^{d-1}dq\sim \xi ^{-1/\nu }  \label{sh}
\end{equation}
If $d>d_{l}$ then this integral converges on small $q$ and this addition to
temperature can be neglected in the RG calculations as it is does not depend
on the correlation length. Note, however that experimentally this correction
is still relevant. Since the values of the fluctuation shift of temperature
are different in the Ising and Lifshitz classes of universality the
temperature of the LP is shifted to lower values then the transition
temperature of the Ising class of universality. The situation of the lower
critical dimension corresponds to logarithmic divergence of $\Delta \tau $,
i.e. $1/\nu =0$. If $d<d_{l}$ then the correction diverges at $\xi
\rightarrow \infty $, which precludes the phase transition at finite
temperatures, the transition temperature goes to zero. According to Equation
(\ref{sh})\ we can consider the exponent $\nu $ in this case to be formally
negative. As we have mentioned above the renormalized value of $d_{l}$\ is
not presently known, so we will consider the two possibilities: $d>d_{l}$
and $d<d_{l}$. Therefore, it does not make sense to use the exponent $\nu $
obtained to order $\varepsilon $, since we do not expect it to be correct
for $d=3$. On the contrary, using the scaling relations (supposed to be
correct even for negative $\nu $) we will express the exponents in the
equation for $\xi (\widetilde{\tau })$ (\ref{first}) via the exponent $\nu $
(generally the exponent $\eta $ is also necessary, however it is zero in the
one-loop approximation). Thus, we arrive at the equation 
\begin{equation}
\left( \widetilde{\tau }+a\widetilde{\lambda }\int_{\xi ^{-1}}^{\infty
}\left( 1+\widetilde{\lambda }q^{-\varepsilon }\right) ^{-1+\left( 4-1/\nu
\right) /\varepsilon }\frac{q^{d-1}dq}{q^{4}}\right) ^{-1}=\xi ^{4}\left( 1+%
\widetilde{\lambda }\xi ^{\varepsilon }\right) ^{-\left( 4-1/\nu \right)
/\varepsilon }  \label{eq}
\end{equation}
Its solution $\xi (\widetilde{\tau })$ substituted into (\ref{sec}) gives $%
\widetilde{\lambda }_{r}(\widetilde{\tau })$ and thus we can obtain the
shift of the LL $\Delta c_{1}(\widetilde{\tau })$ according to (\ref{final}).

The results of numerical evaluation of $\Delta c_{1}(\widetilde{\tau })$
according to Equation (\ref{final}-\ref{eq}) are plotted in Figure 1.\strut
\ We have considered two cases: $d>d_{l}$, $\nu =1$ --- solid line ($%
\widetilde{\lambda }=1$, $a=2$); $d<d_{l}$, $\nu =-1$ --- two dashed curves (%
$\widetilde{\lambda }=1$ for both curves; $a=0.5,a=2$). \ If $d>d_{l}$ then
for all values of parameters $a$ and $\widetilde{\lambda }$ the LL has the
qualitative form as the curve plotted in Figure 1: with decreasing
temperature $\Delta c_{1}(\widetilde{\tau })$ initially increases (due to
increase of $\xi $ as explained above), then for large fluctuations
decreases (due to decrease of the renormalized coupling constant $\lambda
_{r}$) and the LL ends at a Lifshitz point. Note that the LP occurs at the
negative temperatures $\widetilde{\tau }$ due to the \ finite shift of
transition temperature discussed above. For $d<d_{l}$ the situation is more
complex. At large values of $a$ ($a=2$ curve of Figure 1) the LL goes to $%
\widetilde{\tau }\rightarrow -\infty $, so that no Lifshitz point exists (we
may say it is shifted to $\widetilde{\tau }=-\infty $). As can be seen from
Figure 1 in other respects the LL of this case is qualitatively the same as
in the case $d>d_{l}$. At small values of $a$ ($a=0.5$ curve of Figure 1)
two solutions of Equation (\ref{eq}) exist at large temperatures (the one
corresponding to greater $\xi $ is plotted with the dot curve in Figure 1)
and no solution at small temperatures. Thus in this case the LL ends in a
point at which the correlation length is finite, which indicates that this
is a point of the phase transition of the first order. It is not clear
however, whether this case represents physical features or is an artifact of
approximations of the theory.

\strut Comparing obtained results with the experimental data of Schwahn et
al. \cite{Sch99,Sch} we observe that the experimental LL has the same
qualitative shape. We would like to stress that the present study predicts
that the Lifshitz line approaches its mean-field counterpart for high and
deep temperatures, the prediction which is also in agreement with
experimental behavior of the Lifshitz line found in \cite{Sch99,Sch}.
However, on the basis of the current data of Ref. \cite{Sch99,Sch} it is not
clear which of three cases concerning the value of the lower critical
dimension of the system which we have discussed above, actually takes place.

\strut \ Summarizing, we have calculated the deviation of the LL from the
mean-field behavior by taking into account the first fluctuational
correction to the $c_{1}$ coefficient. The renormalized properties are
calculated within the one-loop renormalization group. Our calculations are
carried out in the immediate vicinity of the Lifshitz line, so that no
crossover to either Ising or Brazovsky universality class is considered. The
obtained behavior of the LL qualitatively agrees with that observed in the
scattering experiments on ternary homopolymer/diblock-copolymer systems. The
approach we have presented in the present article is based on the
Landau-Ginzburg expansion of the free energy with phenomenological
parameters $\lambda $, $c_{1}$, $c_{2}$. Of course, this fact restricts a
complete quantitative comparison with experiment. However, we point out that
despite this the agreement of the behavior of the computed Lifshitz line at
high and low temperatures with experimental one is a strong support of the
validity of our approach. The major reason for the uncertainty in the
complete quantitative comparison with the experiment is due to unknown
values of the critical exponents of the Lifshitz universality class.
Depending on the value of the lower critical dimension in the renormalized
theory the LP exists or it is shifted to infinitely small temperatures. More
experiments are needed to determine which of the possibilities actually
takes place, which would amount to experimental determination of the lower
critical dimension of the system. We hope that this work will stimulate such
investigations.

The authors acknowledge stimulating discussions with E. Straube, I.
Erukhimovich, and D. Schwahn. A. K. acknowledges support of the
Graduirtenkolleg ``Polymerwissenschaften''.

\newpage

\section*{Figure Caption}

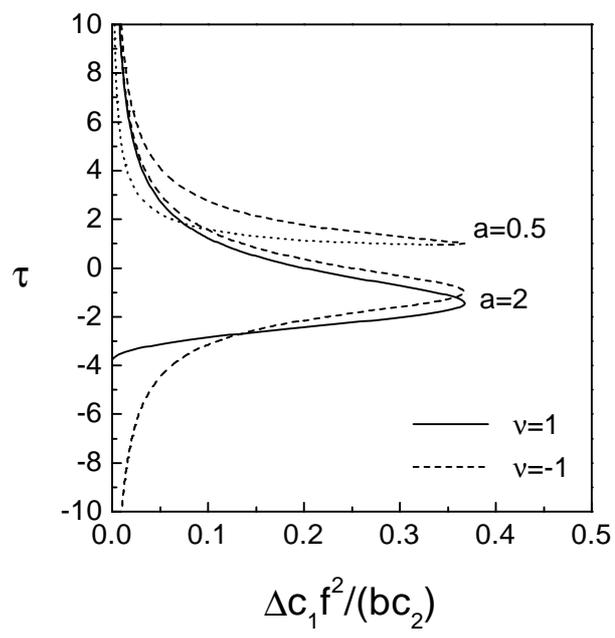
{\bf Figure 1.} The shift of the Lifshitz line$\ \Delta c_{1}$ as a function
of temperature $\widetilde{\tau }$. Solid curve: $d>d_{l}$, dash curves: $%
d<d_{l}$. For all curves $\widetilde{\lambda }=1$.

\end{document}